\documentclass[aps,prl,twocolumn,groupaddress]{revtex4}

\usepackage{amsmath,amssymb,graphicx}
\usepackage{color,times}
\usepackage[latin1]{inputenc}
\usepackage{xcolor}
\usepackage{ulem}
\usepackage{afterpage}
\usepackage[T1]{fontenc}

\definecolor{darkblue}{rgb}{0, 0, 0.8}
\usepackage[colorlinks=true, breaklinks=true, linkcolor=darkblue, citecolor=darkblue, urlcolor=darkblue]{hyperref}
\newcommand{\doilink}[2]{\href{http://dx.doi.org/#1}{#2}}
\newcommand{\beq}{\begin{equation}}
\newcommand{\eeq}{\end{equation}}

\begin{document}

\title{The Collective Lamb Shift of a Nanoscale Atomic Vapour Layer within a Sapphire Cavity}

\author{T. Peyrot}
\author{Y.R.P. Sortais}
\author{A. Browaeys}
\affiliation{Laboratoire Charles Fabry, Institut d'Optique Graduate School, CNRS,
Universit\'e Paris-Saclay, F-91127 Palaiseau Cedex, France}

\author{A. Sargsyan}
\author{D. Sarkisyan}
\affiliation{Institute for Physical Research, National Academy of Sciences - Ashtarak 2, 0203, Armenia}

\author{J. Keaveney}
\author{I.G. Hughes}
\author{C.S. Adams}
\affiliation{Department of Physics, Rochester Building, Durham University,
South Road, Durham DH1 3LE, United Kingdom}

\date{\today}

\begin{abstract}
We measure the near-resonant transmission of light through a dense medium of potassium
vapor confined in a cell with nanometer thickness in order to investigate the origin and validity of
the collective Lamb-shift.
A complete model including the multiple reflections in the nano-cell
accurately reproduces the observed strong asymmetry of the line shape and allows extraction of
a density dependent shift of the atomic resonance.
We observe an additional, unexpected dependence of this shift with the thickness of the
medium. This extra dependence demands further experimental and theoretical investigations.

\end{abstract}

\maketitle
When many light emitters are subjected to an electromagnetic field with a
wavelength $\lambda$, they may react collectively to the field ~\cite{GrossHaroche1983,Guerin2016}.
A well-known example of collective response is the enhancement of the decay rate
of an atomic ensemble with respect to the individual atom case. Owing to the coupling
of atoms via resonant dipole-dipole interactions, it becomes important when the volume of
interaction is smaller than $(\lambda/2\pi)^3$. Collective effects in light scattering have gained a
renewed interest recently with the recognition that they can bias the accuracy of atom-based
sensors such as optical clocks by introducing unwanted energy level shifts\,
\cite{Chang2004,Bromley2016,Campbell2017}.
Alternatively, the collective response can be an asset if properly handled and several recent works suggest how it
can be used to enhance light-matter interfaces\,\cite{Bettles2015,Bettles2016,Shahmoon2017,Perczel2017}.

The resonant dipole-dipole interactions between atoms should lead to a
collective frequency shift of the atomic lines~\cite{Friedberg1973}.
This shift, unfortunately named
the cooperative or collective Lamb-shift (CLS) despite its classical nature, depends on the shape of the sample.
In the case of an atomic slab of thickness $L$ and density $N$ it was
predicted to be~\cite{Friedberg1973}:
\beq\label{Eq:CLS_theo}
{\Delta_{\rm CLS}=\Delta_{\rm LL}-\frac{3}{4}\Delta_{\rm LL}\left(1-{\sin2kL\over2kL}\right)} \ ,
\eeq
where $\Delta_{\rm LL}= -\pi (N/k^3) \Gamma$ is the Lorentz-Lorenz shift, $k=2\pi/\lambda$ is the wave vector
and $\Gamma$ is the natural linewidth of the relevant atomic transition.
Four decades later, the first measurements of the CLS were reported using a layer of Fe
atoms~\cite{Rohlsberger2010} and a slab of hot alkali vapor~\cite{Keaveney2012a}.
Following these experiments it was pointed out that Eq.(\ref{Eq:CLS_theo}) is valid only in
the low density limit ($N/k^3 \ll 1$ with $N$ the density of the vapor~\cite{Javanainen2016,Javanainen2017}), a
condition not met by the experiment  of~\cite{Keaveney2012a} for which $N/k^3\sim 100$.
Reference \cite{Javanainen2014} suggested that this CLS should only be present when large inhomogeneous
broadening is present, such as in hot vapor. However, subsequent experiments on ultracold atoms (insignificant
inhomogeneous broadening) either reported a shift consistent with the CLS prediction~\cite{Roof2016}, or a
negligible shift~\cite{Corman2017}. Recently, theoretical work highlighted that the CLS in a slab
geometry~\cite{Javanainen2016} should merely arise from cavity interferences between the boundaries of the
medium. In contrast to the original suggestion~\cite{Friedberg1973}, in the cavity viewpoint, the CLS  would not be
related to the Lorentz local field. Clearly, the situation is confusing and further work is needed to clarify it.

In this letter, we present a new investigation of the origin and validity of the CLS.
To do so, we measure the transmission resonance line shape of a dense
hot vapor of potassium atoms confined in a slab with nanometer thickness.
We develop a new model to interpret the data based on standard mean-field electromagnetism.
It includes the multiple reflections due to the cavity formed by the two layers of sapphire enclosing the
atomic vapor. We show in particular that Eq.\,(\ref{Eq:CLS_theo}) is valid only in the limit
of a \textit{low-density} atomic slab surrounded by \textit{vacuum}, neither conditions being fulfilled here.
Furthermore, using the model, we deconvolve the cavity effect from the measured transmission
and extract the shift of the atomic
resonance line as a function of density and thickness.
We observe an unexpected oscillatory dependence of the shift with the slab thickness,
which indicates that further refinement of the theory is needed in order to fully account for the optical properties of dense media.

We first give a simple derivation of the CLS [Eq.(\ref{Eq:CLS_theo})] that highlights the roles of
the boundaries and of the dipole-dipole interactions between atoms, as well as its range of applicability.
We consider an atomic slab (thickness $L$, susceptibility $\chi$, refractive index $n=\sqrt{1+\chi}$)
placed in vacuum and illuminated by a plane wave $E_0\exp[ikz]$ with frequency $\omega=ck$.
As the light propagates in the medium, the fields radiated by the induced dipoles
interfere with the incident field and in turn excite new atoms:
the dipole-dipole interaction, which is the interaction of the field radiated by
an atomic dipole with another dipole~\cite{Milonni1996}, is thus included  in the description of the propagation.
The field scattered at position $z$ by a slice of
thickness $dz'\ll \lambda$ located at  position $z'$
is $dE_{\rm sc}(z)=ikP(z')/(2\epsilon_0)\exp[ik|z-z'|] dz'$~\cite{Feynman,Milonni1996}.
Here $P(z')$ is the polarization vector
related to the {\it total} field $E(z')$ inside the medium by $P(z')=\epsilon_0\chi E(z')$.
Consequently, the superposition principle yields the field
transmitted by the slab:
\beq\label{Eq:field_transmitted}
E_{\rm t}(z>L) = E_0 e^{i k z} + {ik\chi\over 2}\int_0^L E(z') e^{ik(z-z')}\, dz'\ .
\eeq
To calculate the {\it total} field inside the slab,
we neglect the multiple reflections at the boundaries between the medium and vacuum.
Therefore $E(z')\approx tE_0e^{inkz'}+rtE_0e^{ink(2L-z')}$, with $n\approx 1+\chi/2$, $t=2/(n+1)\approx 1-\chi/4$ and
$r=(n-1)/(n+1)\approx \chi/4$, for $\chi \ll 1$.
Using these expressions in Eq.~(\ref{Eq:field_transmitted}) we get, up to second order in $\chi$:
\beq\label{Eq:field_transmitted_2}
E_{\rm t}\approx E_0 e^{ikz}\left[1+ i{\chi kL\over 2}
\left(1+ i{ \chi kL\over 4}-{\chi\over 4}+{\chi\over 4}{e^{2ikL}-1\over 2ikL} \right) \right]\ .
\eeq
The susceptibility of the  dilute slab consisting of atoms
with polarizability $\alpha= i(6\pi\Gamma /k^3)/(\Gamma_{\rm t}-2i\Delta)$ ($\Delta=\omega-\omega_0$ with
$\omega_0$ the resonant
frequency, $\Gamma$ the radiative linewidth and $\Gamma_{\rm t}$ the total homogeneous linewidth)
is $\chi = N\alpha$.
Using $1+x\approx 1/(1-x)$ for $|x|\ll 1$ in the parenthesis of
Eq.~(\ref{Eq:field_transmitted_2}), we obtain the transmission coefficient:
\beq
T(\Delta)= \left|{E_{\rm t} \over E_0}\right|^2=
\left|1- {3\pi N L \over k^2}{\Gamma\over \Gamma_{\rm c}-2i(\Delta-\Delta_{\rm c})}\right|^2\ ,
\eeq
with the thickness dependent shift
$\Delta_{\rm c}= -{3\over 4}\Delta_{\rm LL}(1-{\sin2 kL\over 2kL})$
and $\Gamma_{\rm c}=\Gamma_{\rm t}-\frac{3}{4}(kL+{\sin^2 kL\over kL})\Delta_{\rm LL}$.
The offset  $-{3\over 4}\Delta_{\rm LL}$ in the shift is traced back to
the transmission through the first interface.
To recover the extra offset  $\Delta_{\rm LL}$ in Eq.~(\ref{Eq:CLS_theo}),
we must use the Lorentz-Lorenz formula~\cite{Jackson} in Eq.~(\ref{Eq:field_transmitted_2}):
$\chi= N\alpha/(1-N\alpha/3)$.
This derivation therefore shows that
(i) the CLS is a frequency shift of the position of the transmission minimum
and not a shift of the resonance $\omega_0$ of the bulk medium characterized by $\chi$;
(ii) it is a consequence of the reflection of the field at the boundaries of the slab;
(iii) it includes the dipole-dipole interactions in the propagation and,
(iv) Eq.~(\ref{Eq:CLS_theo}) is only valid in a medium for which $\chi\ll 1$ at resonance,
{\it i.e.}  $(N/k^3)(\Gamma/\Gamma_{\rm t})\ll 1$.

To extend the model beyond the dilute regime we include the multiple reflections
in the cavity produced by the interface between the atomic medium and its environment
(index $n_{\rm s}$).
Using a textbook interference argument~\cite{Javanainen2016},
we calculate the transmission coefficient of the field amplitude and get:
\beq\label{Eq:t_amplitude}
t(\Delta)={4 n_{\rm s} n\exp[i(n-n_{\rm s})kL]\over (n_{\rm s}+n)^2-(n_{\rm s}-n)^2 \exp[2inkL]}\ .
\eeq
For $n_{\rm s}=1$, Eq.~(\ref{Eq:t_amplitude})
predicts that the frequency of the minimum  transmission $\Delta_{\rm min}$ does follow
Eq.~(\ref{Eq:CLS_theo}), but only when $(N/k^3)(\Gamma/\Gamma_{\rm t})\ll1$
(see Fig.~4 in~\cite{Javanainen2017}).
This is no longer the case for $n_{\rm s}=1.76$ for which $\Delta_{\rm min}$
{\it never} follows Eq.~(\ref{Eq:CLS_theo}) even at low density (see details in~\cite{SM}).

\begin{figure}
\includegraphics[width=\columnwidth]{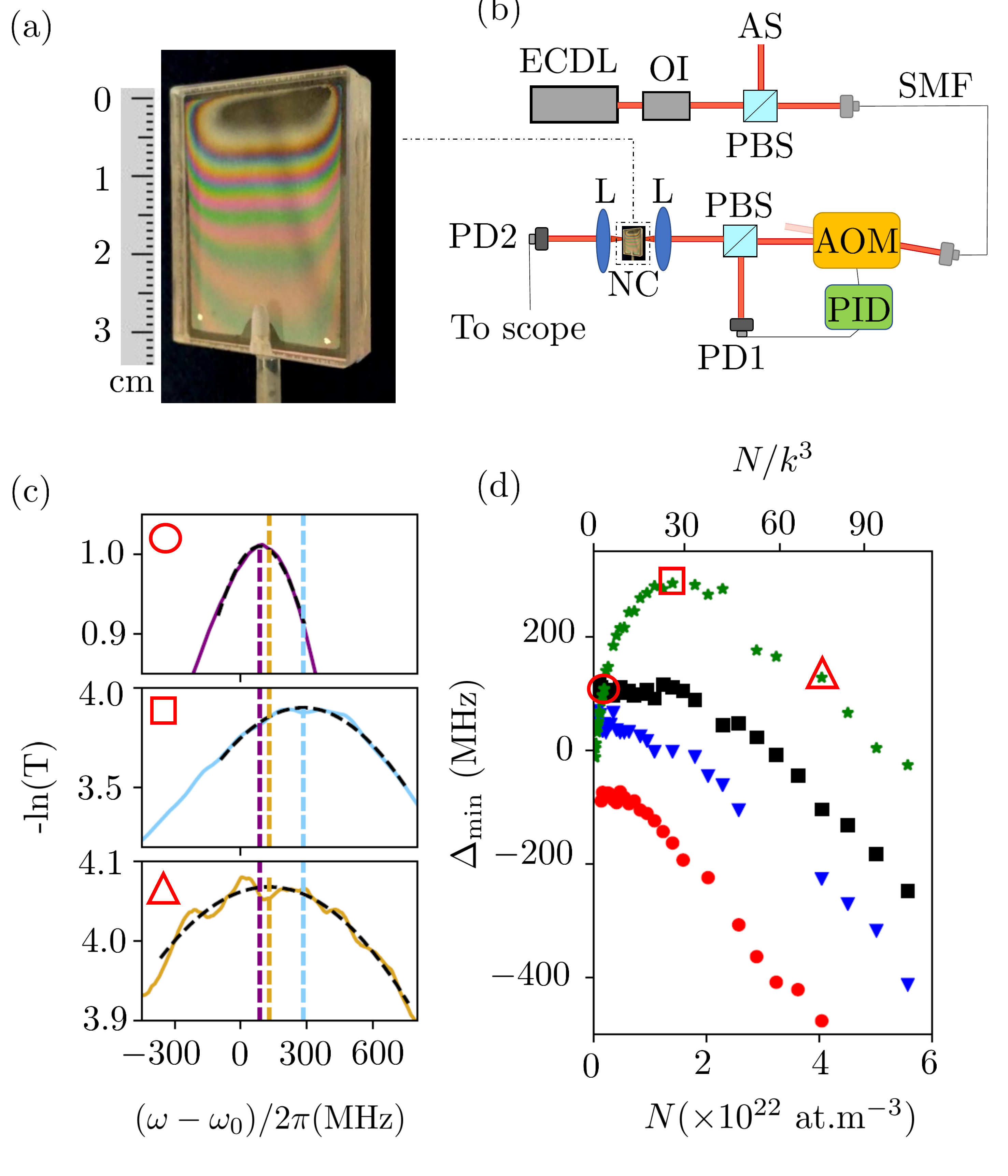}
\caption{(a) Nano-cell used in the experiment. The interference fringes indicate that the thickness of the slab
between the two sapphire windows varies from $50$\,nm to $1.5\,\mu$m at the bottom.
(b) Optical set-up.
ECDL: external cavity laser diode; OI: optical isolator; PBS: polarization beam-splitter;
SMF: single- mode fiber; NC: nano-cell; PD: photodiode;
AOM: acousto-optic modulator; L: lens; AS: absorption spectroscopy;
PID: proportional-integral-derivative controller.
(c) Measured optical density for a slab of thickness $L=490$\,nm as a
function of the detuning and for temperatures
$\Theta=(260^{\circ}{\rm C}, 340^{\circ}{\rm C}, 380^{\circ}{\rm C})$ (top, middle, bottom),
corresponding to $N/k^3 = (3,29,74)$.
Vertical dotted lines: frequency $\Delta_{\rm min}$ of the maximum of the OD.
(d) $\Delta_{\rm min}$ versus density $N$ for
$L = 90$\,nm (red dots),
$L = 110$\,nm (blue triangles), $L = \lambda/2=380$\,nm (black squares)
and $L =3\lambda/4= 575$\,nm (green stars).
The empty square, triangle and  circle correspond to the curves in (c).}
\label{fig1}
\end{figure}

We now describe our experimental investigation of the CLS using a nano-cell~\cite{Sarkisyan2004}.
The nano-cell (Fig.~\ref{fig1}a) consists of two 1 mm-thick sapphire wedge plates ($n_{\rm s}=1.76$)
filled with a vapor of potassium~\cite{Footnote_wedge_plates}.
The resulting atomic slab has a thickness $L$ varying between $50$\,nm and $1.5\,\mu$m ~\cite{Sargsyan2015}.
The atomic density is controlled by heating the cell from room temperature up to $380^{\circ}$C,
achieving similar densities as in~\cite{Keaveney2012a}.
Compared to the earlier measurements performed in  rubidium~\cite{Keaveney2012a}, potassium has the
advantage of a smaller hyperfine splitting in the ground state, which results into a single atomic line at lower densities.
The  optical set-up is shown in Fig.~\ref{fig1}(b).
We measure the transmission of a laser beam
nearly resonant with the D2 transition of $^{39}$K ($\lambda\approx 767$\,nm,  $\Gamma=2\pi\times 6$\,MHz).
The beam is produced by a commercial external cavity laser diode, focused on the
cell sapphire windows with a waist $w\approx40\,\mu$m $\gg L$.
We use the interferometric techniques described in~\cite{Jahier2000}
to measure the local thickness.
The laser is scanned across the resonance over a
range of about $30$\,GHz. The intensity is stabilized using a PID-controlled
acousto-optic modulator~\cite{Luiten2012}. The frequency of the laser is calibrated
by standard saturated absorption spectroscopy in a $7.5$\,cm potassium reference cell.

Figure~\ref{fig1}(c) shows the measured optical density OD, extracted from the transmission
$T$ via ${\rm OD}=-\ln (T)$, as a function of the laser detuning $\Delta$
for three values of the atomic density $N$.
We plot $\Delta_{\rm min}$, defined as the detuning at which the OD is the largest,
as a function of density for various thicknesses $L$ in Fig.~\ref{fig1}(d).
At high density ($N/k^3\gtrsim 20$) we observe a red-shifted, linear variation of $\Delta_{\rm min} $
with $N$ for all $L$.
At low $N$, for  $L>\lambda/2$, $\Delta_{\rm min}$ exhibits a pronounced blue-shift,
and turns into a red-shift at higher density. For thicknesses $L \lesssim \lambda/2$, $\Delta_{\rm min}$
features a plateau at low $N$,
as also seen in~\cite{Keaveney2012a}.
Similar blue-shifts of the minimal transmission were observed  in a nano-cell of cesium~\cite{Maurin2005},
although much smaller than  here, and recently in a slab of ultra-cold rubidium atoms~\cite{Corman2017},
where an evolution from the blue to the red side of the resonance was also measured.

\begin{figure}
\includegraphics[width=\columnwidth]{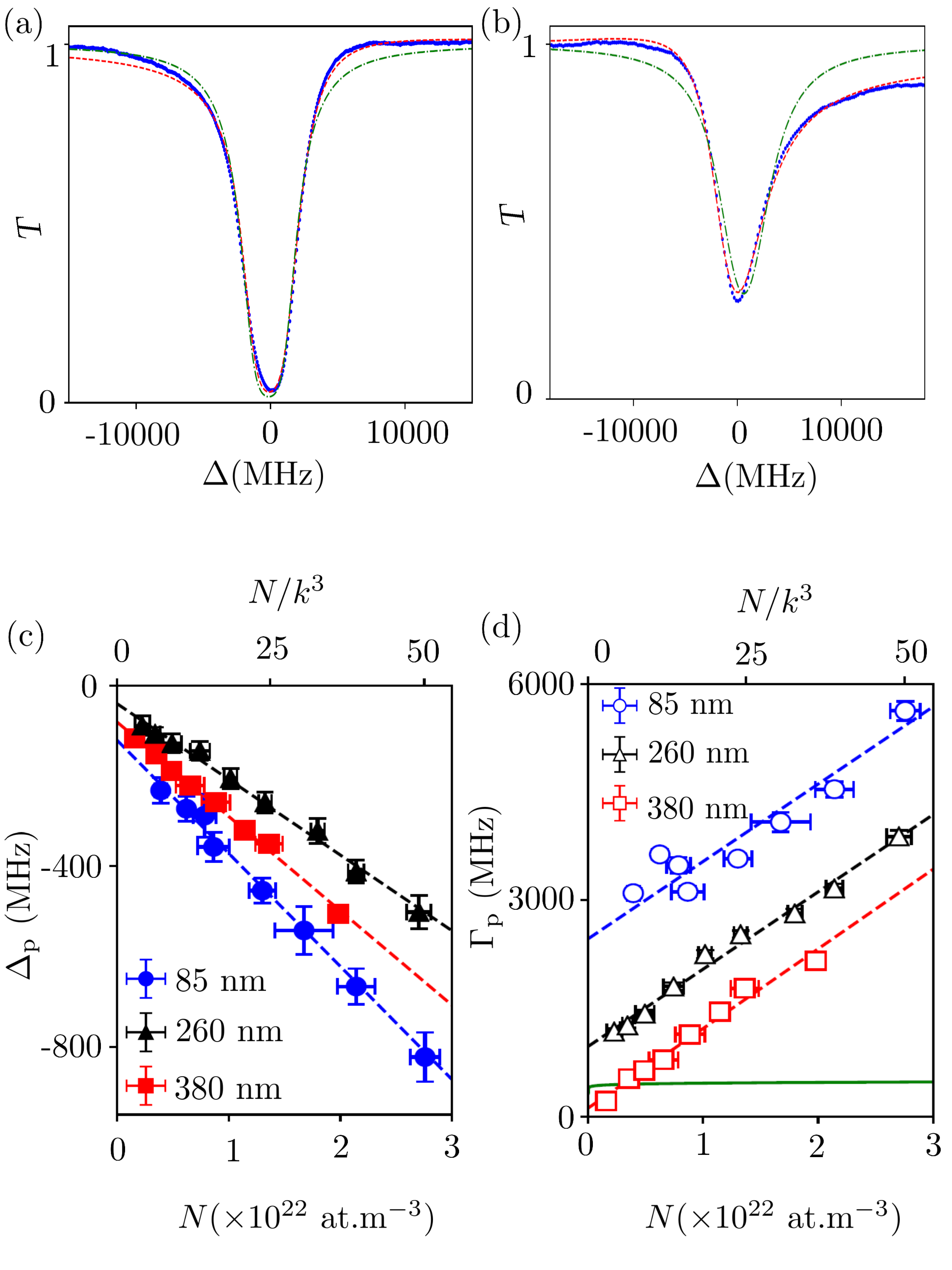}
\caption{Top pannels represent transmission profiles for (a) $\Theta=330^{\circ}$C and $L=440$\,nm and (b)
 $\Theta =365^{\circ}$C at $L=\lambda /4$ where the asymmetry is most pronounced. Blue dots: measured transmission.
Green line: transmission calculated with the model where $n_{\rm s}=1$.
Red dashed line: transmission calculated by the cavity model where $n_{\rm s}=1.76$.
(c) Experimental shift $\Delta_{\rm p}$  and (d)  broadening $\Gamma_{\rm p}$
for various cell thicknesses $L$. Solid line: Doppler width.
The dashed lines are linear fits to the data. The error  bars on both axes are extracted from the fit. }
\label{fig2}
\end{figure}

To explain the data, we now develop a model that deconvolves the effect of the cavity
produced by the interface between the sapphire windows and the atomic medium,
and the bulk  properties of the atomic medium. This was also the approach used in
Ref.~\cite{Keaveney2012a}. However, the model used there to extract the shift
took only partially the cavity effect into account (see details in~\cite{SM}).
Furthermore, as explained above Eq.~(\ref{Eq:CLS_theo}) is irrelevant for
the experimental situation of a nano-cell: the atomic slab should be dilute and surrounded by
vacuum for the formula to hold. The agreement between the measured shift as a function
of the cell thickness and Eq.~(\ref{Eq:CLS_theo}) must therefore be considered as fortuitous.

Our new model incorporates the multiple reflections in the cavity by using Eq.\,(\ref{Eq:t_amplitude}).
As for  the atomic slab, it is described
by a continuous resonant medium with a refractive index $n$.
Ascribing a refractive index to a hot vapor confined in a nano-cell
is far from being obvious, as has been studied
in great details (e.g.~\cite{Schuurmans1976,Vartanyan1995,Zambon1997,Dutier2003a}).
First, the Doppler effect leads to a non-local refractive index, and
second, the small thickness of the cell results in a non-steady state response of most atoms but
the ones flying parallel to the cell surface. However, when the density is as large as
the ones used here, the collisional broadening of the line $\Gamma_{\rm p}$
exceeds the Doppler width $\Delta\omega_{\rm D}$ (see below and Fig.~\ref{fig2}d):
the atomic dipoles
reach their steady-state over a distance $\sim \Delta\omega_{\rm D}/(k\Gamma_{\rm p})$,
much smaller than $L$ and $\lambda$.
It then becomes possible to define a {\it steady-state, local} refractive index~\cite{Vartanyan1995}.

We relate the refractive index of the atomic slab to the electric susceptibility $\chi$
by $n(\Delta)=\sqrt{1+\chi(\Delta)}$. Here we take $\chi=N\alpha_{\rm p}$ with
$\alpha_{\rm p}(\Delta,N)$ the polarizability of the atoms, including the influence of the density
at the single atom level through a broadening and a shift. It is calculated by
summing the contribution of all hyperfine transitions of the D2 line with Lorentzian profiles,
weighted by the corresponding Clebsch-Gordan coefficients~\cite{ElecSus}
($\Gamma$ is the radiative decay rate of the strongest transition):
\beq\label{Eq:susceptibilityELECSUS}
\alpha_{\rm p} (\Delta,N)=
i{6\pi\Gamma\over k^3}
\sum_{F,F'}{C_{FF'}^2\over \Gamma_{\rm t}-2i\Delta_{\rm t}} \ .
\eeq
Here, we do not integrate over the velocity distribution, as Doppler broadening is negligible with respect
to the homogeneous broadening~\cite{Footnote_Doppler}.
In Eq.\,(\ref{Eq:susceptibilityELECSUS}), $\Gamma_{\rm t}=\Gamma +\Gamma _{\rm p}$
is the sum of the radiative linewidth $\Gamma$ and a width $\Gamma_{\rm p}$ that accounts in a phenomenological
way for any broadening mechanism inside the gas beyond the cavity-induced broadening.
In the same way, the detuning $\Delta_{\rm t} = \Delta  +\Delta_{FF'}+\Delta_{\rm p}$,
with $\Delta_{FF'}$ the hyperfine splitting and $\Delta_{\rm p}$ a
phenomenological shift inside the gas beyond the cavity-induced shift.
The quantities $\Delta_{\rm p}(N,L)$ and $\Gamma_{\rm p}(N,L)$ therefore
contain the physics not included in the model:
(i) the interaction of the atoms with the cell walls (only dependent on the
thickness $L$),
(ii)  the collisional dipole-dipole interactions between the light-induced dipoles
(only dependent on the density $N$), and
(iii) any extra effects that may depend both on $L$ and $N$.
Finally, to compare our model to the data, we normalize the
transmission coefficient in intensity to the non-resonant case
($n=1$): $T=|t(\Delta)/ t(\Delta \to \infty)|^2$.

Figures~\ref{fig2}(a-b) show a comparison of the model's prediction
and the measured line shape.
The agreement is very good.
In particular, the model reproduces the observed asymmetric line
shape, and the blue shift of the maximum optical depth
observed in Fig.~\ref{fig1}(d) (see more details in~\cite{SM}). To demonstrate the
importance of the sapphire layers in the optical response, we also plot in Fig.~\ref{fig2}(a) the
result of Eq.~(\ref{Eq:t_amplitude}) for the case of an atomic layer
immersed in vacuum ($n_{\rm s}=1$): there the asymmetry is nearly absent.

To fit the data by the model and obtain the good agreement shown in Figs.~\ref{fig2}(a,b), we
let the density $N$ (or equivalently the temperature $\Theta$~\cite{Footnote_fit_temp}),
the line shift $\Delta_{\rm p}$ and the broadening $\Gamma_{\rm p}$ as free parameters.
In Figs.~\ref{fig2}(c,d) we plot the fitted values of $\Delta_{\rm p}$ and
$\Gamma_{\rm p}$ as a function of the fitted $N$, for various thicknesses.
Both $\Delta_{\rm p}$ and $\Gamma_{\rm p}$
have an offset at asymptotically low density, that increases
when the thickness of the cell decreases. Its origin lies in the interaction
between the atoms and the walls of the nano-cell, as was measured in
Ref.~\cite{Whittaker2014}: when the thickness decreases, the fraction of atoms interacting
significantly with the cell walls increases.
Figure~\ref{fig2}(d) indicates that $\Gamma_{\rm p}$ is much larger than the
Doppler width  and the broadening is dominated by
the density-dependent contribution coming from the collisional dipole-dipole interactions.
For the range of densities explored here, the vapor is thus homogeneously broadened
with $(N/k^3)(\Gamma/\Gamma_{\rm t})\lesssim1$.

\begin{figure}[h]
\includegraphics[width=0.95\columnwidth]{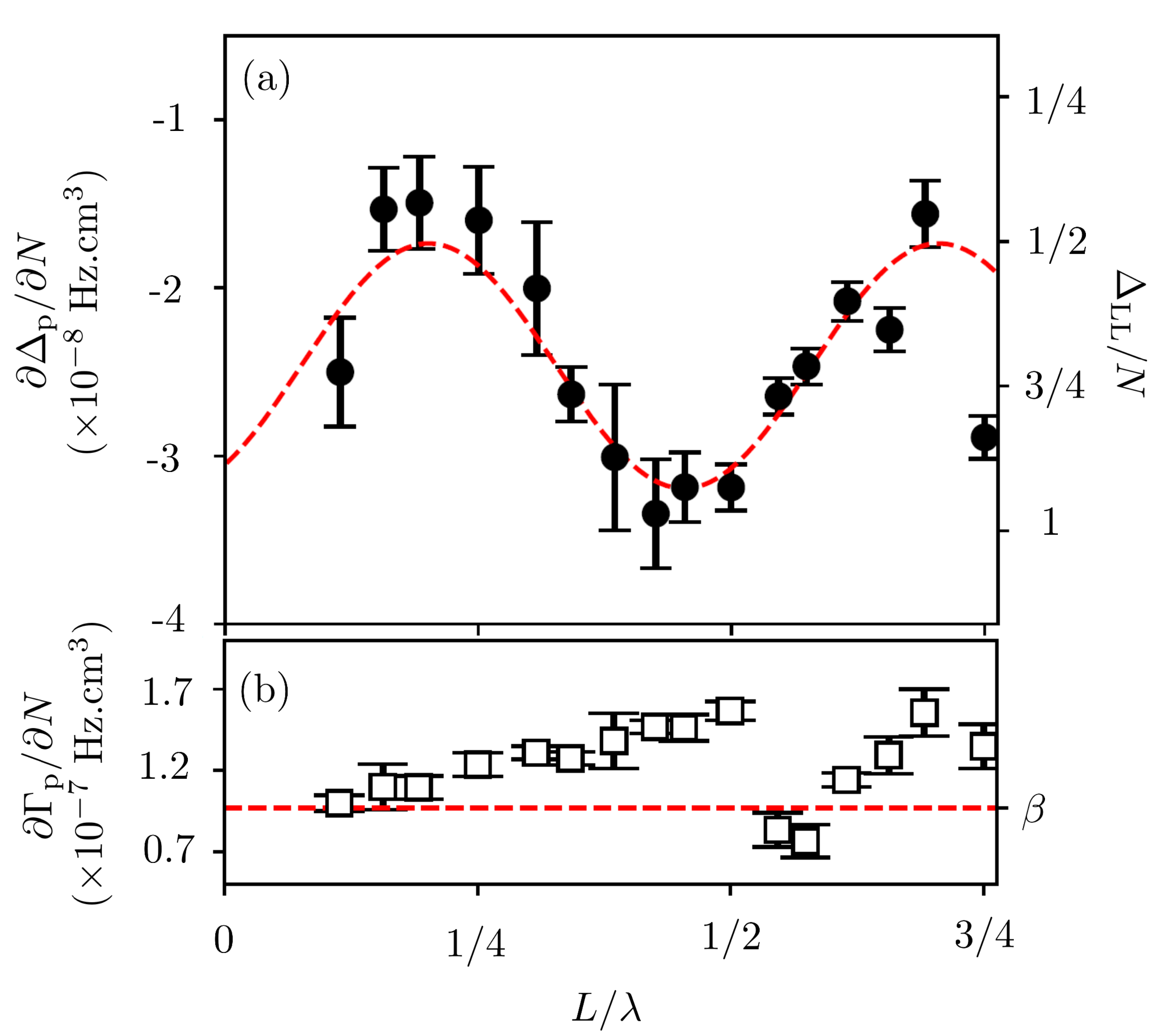}
\caption{(a) Black dots: slope $\partial\Delta_{\rm p}/\partial N$
of the shift extracted from the cavity model
as a function of the cell thickness $L$, together
with the fit by a sinusoidal function (dashed red line). 
(b) Black empty squares: slope $\partial\Gamma_{\rm p}/\partial N$
of the width extracted from the cavity
model  as a function $L$. The red dashed line
is the theoretical value of the self-broadening coefficient $\beta$
resulting from the collisional interactions between atoms (see text).
The errors bars are dominated by the systematic effects detailed in~\cite{SM}.}
\label{fig3}
\end{figure}

To remove the influence of the surface on the shift (an effect depending on $L$ only),
we fit the data presented in Figs.~\ref{fig2}(c,d) by a linear function and extract the
slopes $(\partial\Delta_{\rm p}/\partial N)(L)$
and $(\partial\Gamma_{\rm p}/\partial N)(L)$.
We plot in Figs.~\ref{fig3}(a,b) these slopes as a function of the thickness $L$.
Both quantities feature an offset that we attribute to collisional dipole-dipole interactions between
atoms. For example, the  offset on
$\partial\Gamma_{\rm p}/\partial N$ extracted from Fig.~\ref{fig2}(d) is close to the
calculated self-broadening coefficient resulting from the collisional dipole-dipole interactions
$\beta=2\pi\sqrt{2}\,\Gamma/k^3$~\cite{Weller2011}
(dotted line in Fig.~\ref{fig3}b).
However we also observe a residual oscillation of the shift slope with a
period $(0.5\pm0.02)\lambda$ (the error bars are discussed in ~\cite{SM}).
This oscillation  is unexpected: all known dependences of $\Delta_{\rm p}$
with the cell thickness is included and should result in a shift slope being a bulk property of the medium,
independent of $L$.

We finally examine possible explanations for the residual shift shown in Fig.~\ref{fig3}(a).
A first possibility could be
a non-Maxwellian velocity distribution due to the cell surface~\cite{Todorov2017}.
Dicke coherent narrowing~\cite{Dutier2003b}, which depends on the cell thickness
is also expected in nano-cells.
However the measured line width $\Gamma_{\rm p}$ is
much larger than the expected Doppler width
and any modifications of the velocity distribution
should have a negligible effect on the extracted value of $\Delta_{\rm p}$.
A second possibility could be the influence of the correlations among
dipoles induced by the interactions. They are ignored in our treatment
of the configuration-averaged field and in  all the models developed so
far~\cite{Friedberg1973,Javanainen2014,Javanainen2016,Javanainen2017,Morice1995,Ruostekoski1997}.
This assumption is valid  for dilute gases but it could fail at higher
densities such as the ones explored here. 
Going beyond a mean-field approach
by including them 
could lead to a non-local response of the gas.
The models presented here or in
Refs.~\cite{Friedberg1973,Javanainen2014,Javanainen2016,Javanainen2017},
which assume
a local susceptibility, would then fail -- and including the correlations
would be a highly non-trivial undertaking.

In conclusion, we have performed a new series of measurements of the transmission
of  near-resonant light through an alkali vapor with nanometer scale thickness in order to investigate
the origin and validity of the collective Lamb-shift.
A model, deconvolving the cavity effect from the atomic properties of the slab,
accurately reproduces the observed strong asymmetry of the line shape.
Using this model we extract from our data a shift of the bulk atomic medium
resonance, which oscillates with the thickness of the
medium. The origin of this oscillation is not understood
and we have formulated a few directions that should be explored theoretically.

\begin{acknowledgments}
We acknowledge fruitful discussions with J.~Ruostekoski.
T.~Peyrot is supported by the DGA-DSTL fellowship 2015600028. We also
acknowledge   financial   support   from   EPSRC   (grant
EP/L023024/1) and Durham University.
The data presented  in  this  paper  will be available later on.

\end{acknowledgments}


\begin{thebibliography}{80}

\bibitem{GrossHaroche1983}
M.~Gross and S.~Haroche,
Superradiance: an essay on the theory of collective spontaneous emission,
Phys. Rep. {\bf 93}, 301 (1982).

\bibitem{Guerin2016}
W.~Gu\'erin, M.T.~Rouabah and R.~Kaiser,
Light interacting with atomic ensembles: collective, cooperative and mesoscopic effects,
\doilink{10.1080/09500340.2016.1215564}{J. Mod. Optics {\bf 64},  895 (2017)}.

\bibitem{Chang2004}
D.E.~Chang, J.~Ye and M.D.~Lukin,
Controlling dipole-dipole frequency shifts in a lattice-based optical atomic clock,
\doilink{10.1103/PhysRevA.69.023810}{Phys. Rev. A {\bf 69}, 023810 (2004)}

\bibitem{Bromley2016}
S.L.~Bromley {\it et al.},
Collective atomic scattering and motional effects in a dense coherent medium,
\doilink{10.1038/ncomms11039}{Nat. Comm. {\bf 7}, 11039 (2016)}.

\bibitem{Campbell2017}
S.L.~Campbell {\it et al.},
A Fermi-degenerate three-dimensional optical lattice clock,
\doilink{10.1126/science.aam5538}{Science. {\bf 358}, 6359 (2017)}

\bibitem{Bettles2015}
R.J.~Bettles, S.A.~Gardiner and C.S.~Adams,
Cooperative ordering in lattices of interacting two-level dipoles,
\doilink{10.1103/PhysRevA.92.063822}{Phys. Rev. A {\bf 92}, 063822 (2015)}.

\bibitem{Bettles2016}
R.J.~Bettles, S.A.~Gardiner and C.S.~Adams,
Enhanced optical cross cection via collective coupling of atomic dipoles in a 2D array,
\doilink{10.1103/PhysRevLett.116.103602}{Phys. Rev. Lett. {\bf 116}, 103602 (2016)}.

\bibitem{Shahmoon2017}
E.~Shahmoon, D.S.~Wild, M.D.~Lukin and S.F.~Yelin,
Cooperative resonances in light scattering from two-dimensional atomic arrays,
\doilink{10.1103/PhysRevLett.118.113601}{Phys. Rev. Lett. {\bf 118}, 113601 (2017)}

\bibitem{Perczel2017}
J.~Perczel {\it et al.},
Topological quantum optics in two-dimensional atomic arrays,
\doilink{https://doi.org/10.1103/PhysRevLett.119.023603}{Phys. Rev. Lett. {\bf 119}, 023603 (2017)}

\bibitem{Friedberg1973}
R.~Friedberg, S.R.~Hartmann and J.T.~Manassah,
Frequency shift and absorption by resonant systems of two-level atoms,
Phys. Rep. {\bf 7}, 101 (1973).

\bibitem{Rohlsberger2010}
R.~R\"ohlsberger, K.~Schlage, B.~Sahoo, S.~Couet and R.~Roeffer,
Collective Lamb shift in single-photon superradiance,
\doilink{10.1126/science.1187770}{Science {\bf 328}, 1248 (2010)}.

\bibitem{Keaveney2012a}
J.~Keaveney, A.~Sargsyan, U.~Krohn, I.G.~Hughes, D.~Sarkisyan and C.S.~Adams,
Cooperative Lamb shift in an atomic vapor layer of nanometer thickness,
\doilink{10.1103/PhysRevLett.108.173601}{Phys. Rev. Lett. {\bf 108}, 173601 (2012)}.		

\bibitem{Javanainen2016}
J.~Javanainen and J.~Ruostekoski,
Light propagation beyond the mean-field theory of standard optics,
\doilink{10.1364/OE.24.000993}{Optics Express,  {\bf 24}, 993 (2016)}.

\bibitem{Javanainen2017}
J.~Javanainen, J.~Ruostekoski, Y. Li and S.-M.~Yoo,
Exact electrodynamics versus standard optics for a slab of cold dense gas,
\doilink{10.1103/PhysRevA.96.033835}{Phys. Rev. A {\bf 96}, 033835 (2017)}

\bibitem{Javanainen2014}
J.~Javanainen, J.~Ruostekoski, Y.~Li and S.-M.~Yoo,
Shifts of a resonance line in a dense atomic sample,
\doilink{10.1103/PhysRevLett.112.113603}{Phys. Rev. Lett. {\bf 112}, 113603 (2014)}.

\bibitem{Roof2016}
S.J.~Roof, K.J.~Kemp, M.D.~Havey and I.M.~Sokolov,
Observation of single-photon superradiance and the cooperative Lamb shift in an extended sample of cold atoms,
\doilink{10.1103/PhysRevLett.117.073003}{Phys. Rev. Lett. {\bf 117}, 073003 (2016)}.

\bibitem{Corman2017}
L.~Corman, J.-L.~Ville, R.~Saint-Jalm, M.~Aidelsburger, T.~Bienaimé, S.~Nascimbène, J.~Dalibard and J.~Beugnon,
Transmission of near-resonant light through a dense slab of cold atoms,
\doilink{10.1103/PhysRevA.96.053629}{Phys. Rev. A \textbf{96}, 053629 (2017)}.

\bibitem{Milonni1996}
H. Fearn, D.F.V James, and P. Milonni,
Microscopic approach to reflection, transmission and the Ewald-Osen extinction theorem,
\doilink{10.1119/1.18315}{Am. J. Phys. {\bf 64}, 986 (1996).}

\bibitem{Feynman}
R.P. Feynman, R.B. Leighton, and M. Sands,
{\it Lectures on Physics}, vol. 1, chap. 30,
Addison Wesley (2006).

\bibitem{Jackson}
J.D.~Jackson,
{\it Classical Electrodynamics},
(John Wiley and Sons, New York, 1998).

\bibitem{SM} See Supplemental Material.

\bibitem{Sarkisyan2004}
D.~Sarkisyan, T.~Varzhapetyan, A.~Sarkisyan, Y.~Malakyan, A.~Papoyan, A.~Lezama, D.~Bloch and M.~Ducloy,
Spectroscopy in an extremely thin vapor cell: Comparing the cell-length dependence in fluorescence and in absorption techniques,
\doilink{10.1103/PhysRevA.69.065802}{Phys. Rev. A {\bf 69}, 065802 (2004).}

\bibitem{Footnote_wedge_plates} To avoid cavity effect from the sapphire plates, they form  a
wedge with an angle of $10$\,mrad, much larger than the $0.1$\,mrad angle of the atomic slab.

\bibitem{Sargsyan2015}
A.~Sargsyan, A.~Tonoyan, {\it et al.},
Complete hyperfine Paschen-Back regime at relatively small magnetic fields realized in potassium nano-cell.
\doilink{10.1209/0295-5075/110/23001}{Europhysics Letters {\bf 110},  23001 (2015).}

\bibitem{Jahier2000}
E.~Jahier, J.~Gu\'ena, P.~Jacquier, M.~Lintz, A.V.~Papoyan and M.A.~Bouchiat,
Temperature-tunable sapphire windows for reflection loss-free operation of vapor cells,
\doilink{10.1007/s003400000388}{Appl. Phys. B {\bf 71}, 561 (2000).}

\bibitem{Luiten2012}
G.~Truong, J.D.~Anstie, E.F.~May, T.M.~Stace and A.N~Luiten,
Absolute absorption line-shape measurements at the shot-noise limit,
\doilink{10.1103/PhysRevA.86.030501}{Phys. Rev. A {\bf 86}, 030501 (2012)}

\bibitem{Maurin2005}
I.~Maurin, P.~Todorov, I.~Hamdi, A.~Yarovitski, G.~Dutier, D.~Sarkisyan, S.~Saltiel, M.-P.~Gorza, M.~Fichet, D.~Bloch and M.~Ducloy,
Probing an atomic gas confined in a nanocell,
\doilink{10.1088/1742-6596/19/1/003}{J. Physics: Conference Series {\bf 19}, 20 (2005)}.

\bibitem{Schuurmans1976}
M.F.H. Schuurmans,
Spectral narrowing of selective reflection,
\doilink{10.1051/jphys:01976003705046900}{J. Phys. France {\bf 37}, 469 (1976)}.

\bibitem{Vartanyan1995}
T.A. Vartanyan and D.L. Lin,
Enhanced selective reflection from a thin layer of a dilute gaseous medium,
\doilink{10.1103/PhysRevA.51.1959}{Phys. Rev. A {\bf 51}, 1959 (1995)}.

\bibitem{Zambon1997}
B. Zambon and G. Nienhuis,
Reflection and transmission of light by thin vapor layers,
\doilink{10.1016/S0030-4018(97)00331-3}{Opt. Comm. {\bf 143}, 308 (1997)}.

\bibitem{Dutier2003a}
G. Dutier, S. Saltiel, D. Bloch, and M. Ducloy,
Revisiting optical spectroscopy in a thin vapor cell: mixing of reflection and transmission as a Fabry-Perot microcavity effect,
\doilink{10.1364/JOSAB.20.000793}{JOSA B {\bf 20}, 793 (2003)}.

\bibitem {ElecSus}
M.A.~Zentile, J.~Keaveney, L.~Weller, D.J.~Whiting, C.S.~Adams and I.G.~Hughes,
ElecSus: A program to calculate the electric susceptibility of an atomic ensemble,
\doilink{10.1016/j.cpc.2014.11.023}{Comp. Phys. Comm. {\bf 189}, 162 (2015)}.

\bibitem{Footnote_Doppler} By performing the integration over the velocity distribution,
we have checked that it indeed has no effect.

\bibitem{Footnote_fit_temp} Although the temperature is left as a free parameter
here, the fitted values are very close to the values measured on the experiment by the thermocouple in
contact with the reservoir.

\bibitem{Whittaker2014}
K.A.~Whittaker, J.~Keaveney, I.G.Hughes, A.~Sargsyan, D.~Sarkisyan and C.S.~Adams,
Optical response of gas-phase atoms at less than $\lambda/80$ from a dielectric surface,
\doilink{10.1103/PhysRevLett.112.253201}{Phys. Rev. Lett. {\bf 112}, 253201 (2014)}.

\bibitem{Weller2011}
L.~Weller, R.J.~Bettles, P.~Siddons, C.S.~Adams and I.G.~Hughes,
Absolute absorption on the rubidium D1 line including resonant dipole-dipole interactions,
\doilink{10.1088/0953-4075/44/19/195006}{J. Phys. B {\bf44}, 195006 (2011)}.

\bibitem{Todorov2017}
P. Todorov, D. Bloch,
Testing the limits of the Maxwell distribution of velocities for atoms flying nearly parallel to the walls of a thin cell,
\doilink{10.1063/1.4997566}{The Journal of Chemical Physics  {\bf 147}, 194202 (2017)}.

\bibitem{Dutier2003b}
G. Dutier, A. Yarovitski, S. Saltiel, A. Papoyan, D. Sarkisyan, D. Bloch and M. Ducloy,
Collapse and revival of a Dicke-type coherent narrowing in a sub-micron thick vapor cell transmission spectroscopy,
Europhysics Letters {\bf 63}, 35 (2003).


\bibitem{Morice1995}
O.~Morice, Y.~Castin and J.~Dalibard,
Refractive index of a dilute Bose gas,
\doilink{10.1103/PhysRevA.51.3896}{Phys. Rev. A {\bf 51}, 3896 (1995)}.

\bibitem{Ruostekoski1997}
J.~Ruostekoski and J.~Javanainen,
Quantum field theory of cooperative atom response: Low light intensity,
\doilink{10.1103/PhysRevA.55.513}{Phys. Rev. A {\bf 55}, 513 (1997)}.


\end{thebibliography}
\end{document}


\title{Supplemental Material: The Collective Lamb Shift of a Nanoscale Atomic Vapour Layer within a Sapphire Cavity}

\author{T. Peyrot}
\author{Y.R.P. Sortais}
\author{A. Browaeys}
\affiliation{Laboratoire Charles Fabry, Institut d'Optique Graduate School, CNRS,
Universit\'e Paris-Saclay, F-91127 Palaiseau Cedex, France}

\author{A. Sargsyan}
\author{D. Sarkisyan}
\affiliation{2 Institute for Physical Research, National Academy of Sciences - Ashtarak 2, 0203, Armenia}

\author{J. Keaveney}
\author{I.G. Hughes}
\author{C.S. Adams}
\affiliation{Department of Physics, Rochester Building, Durham University,
South Road, Durham DH1 3LE, United Kingdom}

\date{\today}

\maketitle

In this Supplemental Material,  we first provide more informations about the model used in
Ref.~\cite{Keaveney2012a} to extract the Collective Lamb Shift in a nanocell of rubidium.
We then show that the cavity model introduced in the main text reproduces the transition
from the blue to the red side of the atomic resonance.
In the third section, we compare the situation of an atomic slab placed in vacuum to the case where it is
placed between two sapphire plates.
In the last section, we explain how the error bars displayed in Figs.~3(a,b) of the main text are calculated.

\section{Model used in Ref.~\cite{Keaveney2012a} to extract the CLS shift}

The model developed in \cite{Keaveney2012a} relies on two ingredients. 
First, a model to calculate the susceptibility $\chi$ of the atomic slab, and therefore its refractive index. 
This model is the same as the one we used in the present work, with the same free parameters $\Delta_{\rm p}$ and 
$\Gamma_{\rm p}$ and density $N$.
Second, a model to include the cavity surrounding the slab, which we now describe. 

The cavity model used the following formula for the transmission
of the atomic slab (refractive index $n$) contained between the two sapphire plates
(refractive index $n_s$):
\begin{equation}\label{Eq:K2012}
T_r(\Delta)= T(\Delta)\ {1-R[n(\Delta)]\over 1-R[n=1]}\ .
\end{equation}
Here, $T(\Delta)$ was assumed to follow a Beer-Lambert attenuation: $T(\Delta)=e^{-2n''(\Delta)kL}$ where $k=2\pi/\lambda$ is the wave vector of light in vacuum, $L$ is the cell thickness and $n''={\rm Im}[\sqrt{1+\chi}]$ with $\chi$ the susceptibility of the atomic vapor. The susceptibility was calculated
using the multilevel model described in the main text.
The reflection coefficient in intensity $R[n(\Delta)]$ was calculated using the formalism
developed in Ref.~\cite{Brooker}:
\begin{equation}\label{Eq:Brooker}
R[n(\Delta)]=\left|{Z_{\rm in}-Z_s\over Z_{\rm in}+Z_s}\right|^2\ ,
\end{equation}
with
\begin{equation}
Z_{\rm in}=\frac{Z_s-iZ_n\tan(kL)}{1-i(Z_s/Z_n)\tan(kL)}\ .
\end{equation}
In this last equation, $Z_n=1/n(\Delta)$, with $n=\sqrt{1+\chi}$, and $Z_s=1/n_s$.
The factor $1-R[n(\Delta)]$ was phenomenologically introduced to
account for the light reflected by the cavity formed by the
two sapphire plates.
When using Eq.\eqref{Eq:Brooker}, Ref.~\cite{Keaveney2012a} assumed that
the refractive index of the atomic slab was equal to 1.
Although this approximation is very good for dilute system, it is not valid
close to the resonance for the densities reached in the experiment,
where ${\rm Re}[n]$ can be as high as 1.3~\cite{Keaveney2012b}.
Consequently the approximation made in \cite{Keaveney2012a} did not
include the cavity effects in an accurate way.

Using this cavity model, Ref.~\cite{Keaveney2012a}, fitted the data to extract 
$\Delta_{\rm p}$ and  $\Gamma_{\rm p}$, as we did in the present work. The authors
plotted the gradient $\partial\Delta_{\rm p}/\partial N$ as a function of the cell thickness $L$
and found that it agreed with the CLS formula [Eq.\,(1) of main text]. However,
considering the facts that 
(i) the cavity effect being independently taken into account, $\Delta_{\rm p}$ cannot be compared 
to the CLS, which only originates from the cavity,
(ii) the cavity model used to extract $\Delta_{\rm p}$ was not accurately taken into account, and
(iii) the CLS formula  {\it does not} apply to an
atomic slab placed between two sapphire plates (see also Sec.~\ref{Sec:compareCLS}),
the agreement between the shift measured in ~\cite{Keaveney2012b} and the CLS formula
must be considered fortuitous.
We note here that if we apply the model described in this section 
(used in Ref.~\cite{Keaveney2012a}) to our new potassium data, 
we find the same dependence of the shift slope $\partial\Delta_{\rm p}/\partial N$ with $L$
as the one obtained in Ref.~\cite{Keaveney2012a} using rubidium. 
This indicates that our new measurements are compatible with the ones on rubidium. 

\section{Transition from blue to red shift in the cavity model }\label{Sec:Delta_min}

\begin{figure}
\includegraphics[width=0.6\columnwidth]{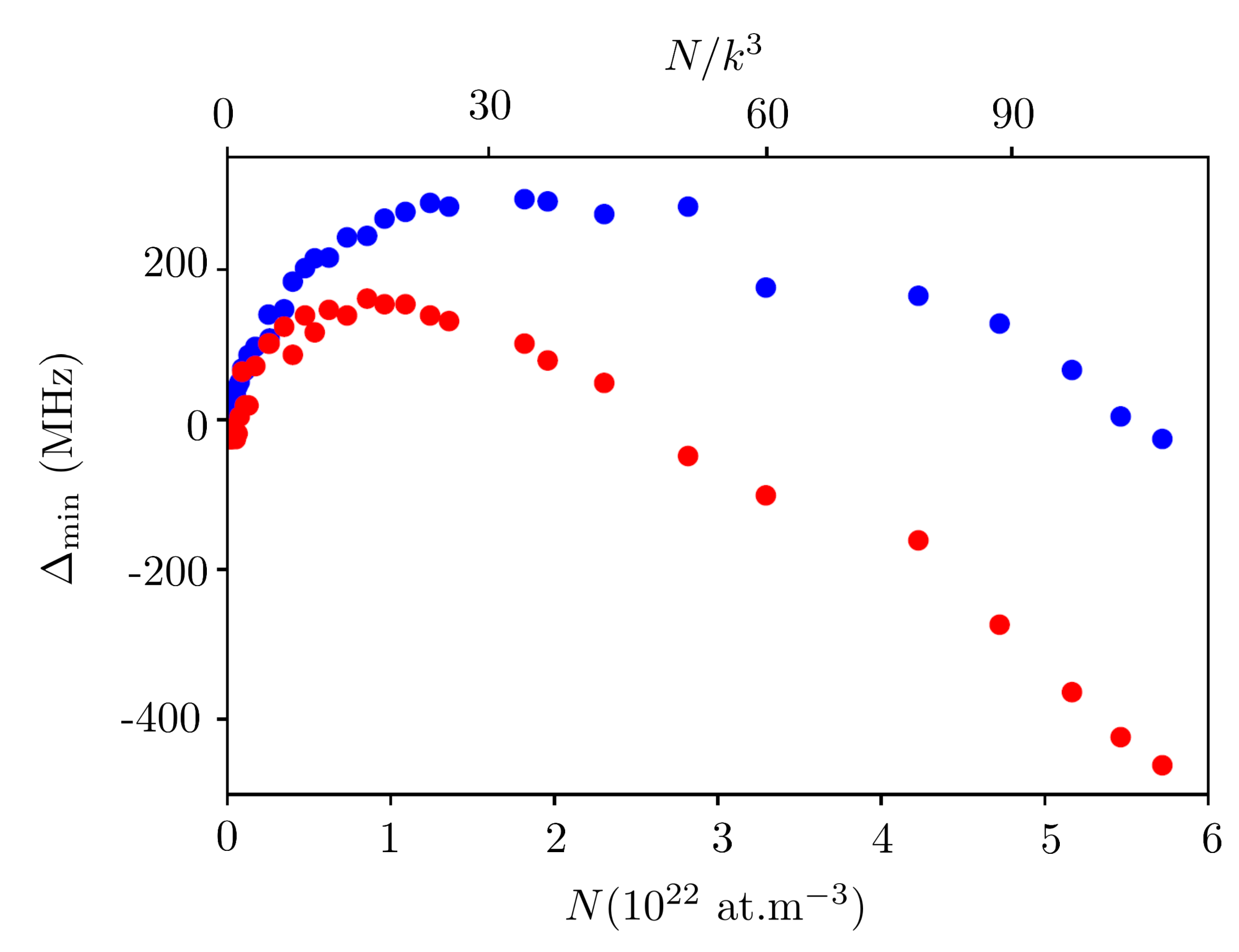}
\caption{Value of $\Delta_{\rm min}$ for $L=490$\,nm as a function of $N$.
Blue dots: experimental values. Red dots: value from the best theoretical fit by the cavity model.}
\label{fig_Deltamin}
\end{figure}

In this second Section we show that the cavity model introduced in the main text
reproduces the transition of
the frequency $\Delta_{\rm min}$ of the largest optical depth from
the blue to the red side of the resonance.

For each measured transmission spectra, we fit the data by the cavity model and extract the
three parameters $(N,\Gamma_{\rm p},\Delta_{\rm p})$. We then calculate
from these best theoretical spectra the frequency detunings $\Delta_{\rm min}$
corresponding to  the largest optical depths. Examples of results are
plotted in Fig.~\ref{fig_Deltamin} for a cell thickness $L=490$\,nm, together with the measured
values. We observe that the transition from blue to red with increasing density is qualitatively
reproduced by the cavity model. However, there is no quantitative agreement despite the fact
the cavity model does reproduce well the measured line shape. The discrepancy between
measured and calculated values of $\Delta_{\rm min}$ is nonetheless smaller than a tenth
of the linewidth of the line, making it sensitive to possible systematic errors, such as the
normalization procedure that we detail in Section~\ref{Sec:errorbars}.
However, as we show in Section~\ref{Sec:errorbars}, the oscillations observed in $\Delta_p$ in Fig.~3(a)
are robust against these systematics.

Finally, one way to understand qualitatively the asymmetry of the line and 
the blue shift observed at low densities uses the  complex susceptibility $\chi$ of the vapor. 
It has a Lorentzian profile and is proportional to $N$. 
When $N$ increases the imaginary part $n''$ of the index of refraction 
$n=\sqrt{1+\chi}$, is a combination of the real and imaginary parts of $\chi$. This is enough
to lead to an asymmetry of the line and a blue shift of $n''$, and hence of the transmission line.

\section{Comparison of an atomic slab in vacuum and between two sapphire plates}\label{Sec:compareCLS}

\begin{figure}
\includegraphics[width=\columnwidth]{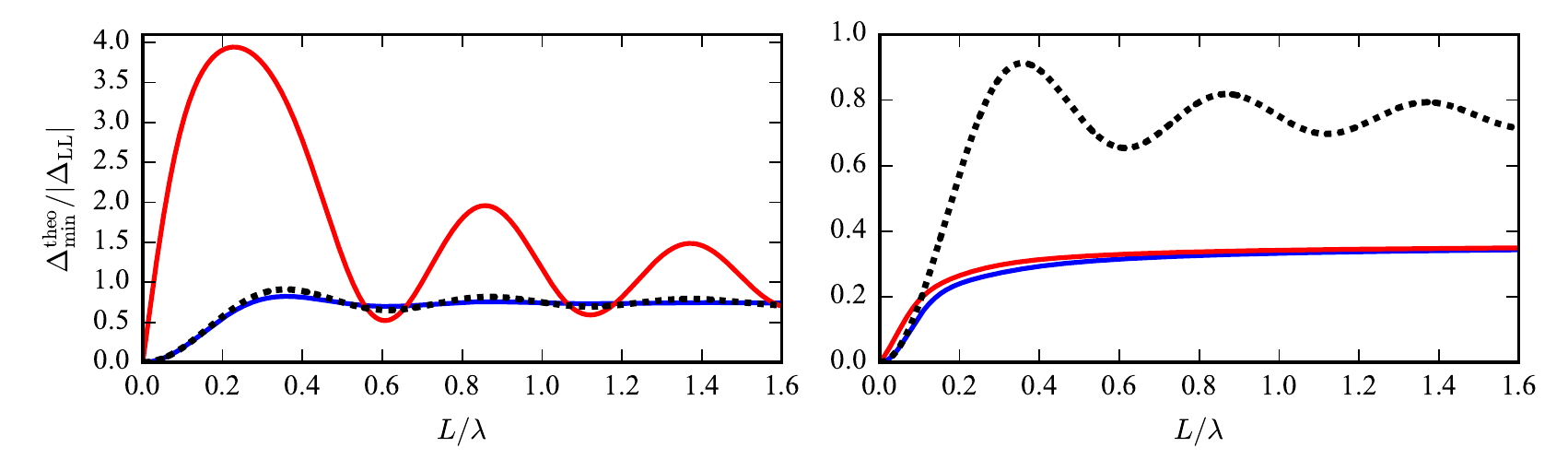}
\caption{Frequency $\Delta_{\rm min}^{\rm theo}/|\Delta_{\rm LL}|$ of the minimum of transmission
as a function of the cell thickness for two values of the density corresponding
to $(N/k^3)(\Gamma/\Gamma_{\rm t})=0.008$ (left) and $0.2$ (right). The black dashed line is the prediction of the thickness
dependent part of the CLS [Eq.(1) of main text]. The solid lines correspond to the predictions of the cavity model
developed in the main text for a slab in vacuum ($n_{\rm s}=1$, blue solid line) or between two sapphire plates
($n_{\rm s}=1.76$, red solid line).}
\label{Fig_SM_CompareDeltaMinshift}
\end{figure}

In this third section, we compare the situation of an atomic slab placed in vacuum or between
two plates of refractive index $n_{\rm s}$, as is the case
for a nano-cell.
We model the atomic vapor  by a continuous medium with a susceptibility given by $\chi=N\alpha$,
where  $\alpha= i(6\pi\Gamma /k^3)/(\Gamma_{\rm t}-2i\Delta)$  is the polarizability of the atoms, and a refractive index
$n=\sqrt{1+\chi}$.

We use Eq.\,(5) of the main text for the transmission coefficient.
We calculate the frequency $\Delta_{\rm min}^{\rm theo}/|\Delta_{\rm LL}|$ corresponding to the minimum of transmission
for different values of the density and cell thickness.
We plot in Fig.~\ref{Fig_SM_CompareDeltaMinshift} the results for two densities corresponding
to $(N/k^3)(\Gamma/\Gamma_{\rm t})=0.01$ and $0.2$, for the case of a slab immersed in
vacuum ($n_{\rm s}=1$) or placed between two sapphire plates ($n_{\rm s}=1.76$).
We also plot the prediction of the thickness dependent
CLS for $\chi=N\alpha$ (see main text):
$\Delta_{\rm c}= -{3\over 4}\Delta_{\rm LL}\left(1-{\sin2 kL\over 2kL}\right)$.
Note that taking for the susceptibility the Lorentz-Lorenz expression $\chi=N\alpha/(1-N\alpha/3)$, would simply offset all predictions of $\Delta_{\rm min}^{\rm theo}/|\Delta_{\rm LL}|$ in fig.~\ref{Fig_SM_CompareDeltaMinshift} by $-1$; in particular we would recover the prediction of Eq.(1) of the main text for the CLS.

We observe that at low density and for a slab immersed in vacuum,
the frequency of the minimum of transmission follows the prediction of the CLS formula
(the agreement is perfect at asymptotically low density). However, when the slab is confined between
sapphire plates, this is no longer the case: there the multiple reflections induced
by the cavity cannot be neglected, as is the case to derive the CLS formula (see main text).
At higher density, the frequency of the minimum of transmission deviates from the
prediction of the CLS formula. The frequency $\Delta_{\rm min}^{\rm theo}$ becomes
independent of the thickness of the cell
when the density increases and is nearly identical for the case of vacuum and sapphire.

Experimentally, however, it is not possible to check directly this prediction by studying the
frequency of the transmission minimum
$\Delta_{\rm min}^{\rm exp}$:
$\Delta_{\rm min}^{\rm exp}$ also depends on the cell thickness via the surface interaction (offset
of $\Delta_{\rm p}$ in Fig.2c of main text), adding an extra dependence with $L$.

\section{Discussion of systematic errors and calculation of the error bars  on
$\partial \Delta_{\rm p}/\partial N$ and $\partial \Gamma_{\rm p}/\partial N$}\label{Sec:errorbars}

In this last Section we explain how we calculate the error bars on the slopes
$\partial \Delta_{\rm p}/\partial N$ and $\partial \Gamma_{\rm p}/\partial N$
shown in Figs.~3(a,b) of the main text. We show in particular that the oscillations
observed in Fig.~3(a) are robust against possible systematic effects.
This discussion is necessary for two reasons. Firstly, as we saw in Sec.~\ref{Sec:Delta_min},
despite the fact that the cavity model reproduces very well the lineshape of the transmission
signal, it fails to yield the measured value of $\Delta_{\rm min}$.
The difference between the measured values and the values extracted from
the cavity model is on the same order than the values of the shift of interest $\Delta_{\rm p}$
obtained by a fit of the line shape with a very small error bar. This indicates that the error bars
from the fit possibly underestimate systematic effects. Secondly, as pointed out in the main text,
we stabilized the intensity of the laser during the scan. However a small residual variation
is possible, that would lead to an enhancement or a reduction of the asymetry of the line,
thus translating into a possible bias on the fitted value of $\Delta_{\rm p}$.

To test the influence of a residual variation of intensity during the scan
on the result of the fit yielding $\Delta_{\rm p}$ and  $\Gamma_{\rm p}$, we
study two different ways of normalizing the data:
(1) normalization by a linear function interpolating the first and last points of the spectrum,
(2) normalization by the first value of the transmission of the spectrum, which therefore
yields $T=1$.
We fit the lines, normalized by the two procedures, by the cavity model
and extract $\Delta_{\rm p}$ and $\Gamma_{\rm p}$, together with their error bars.
We finally extract the slopes
$\partial \Delta_{\rm p}/\partial N$ and $\partial \Gamma_{\rm p}/\partial N$, together with
the error bars from the fit.

\begin{figure}
\includegraphics[width=0.6\columnwidth]{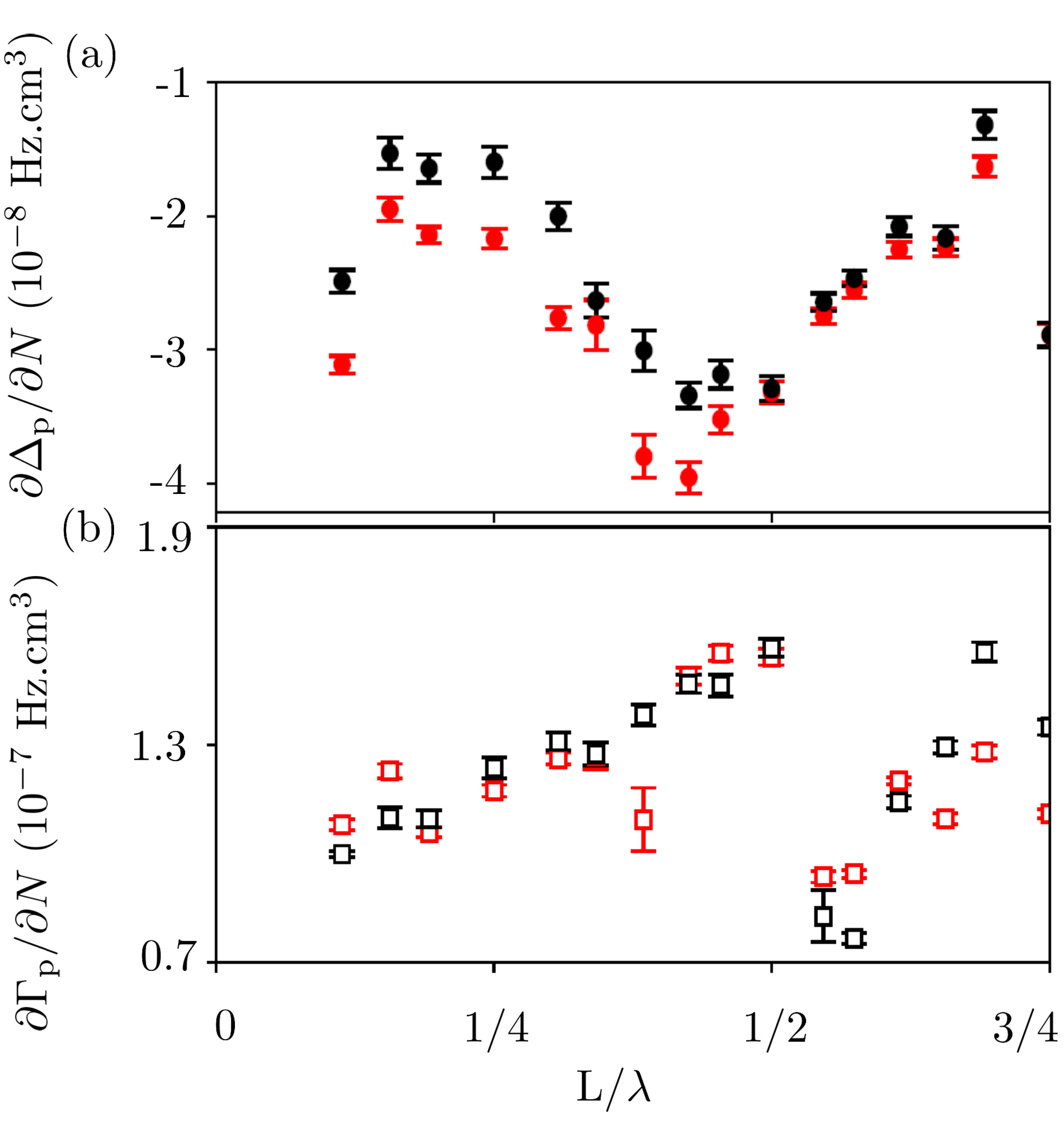}
\caption{
(a) Slope of the shift $(\partial \Delta_{\rm p}/\partial N)(L)$
extracted from the cavity model as a function of the cell thickess $L$.
(b) Slope of the width $(\partial \Gamma_{\rm p}/\partial N)(L)$ from the cavity model.
(All figures) Red markers: normalization procedure (1), see text.
Black markers: normalization procedure (2).
Here, the error bars are given by the fit. }\label{fig3}
\end{figure}

Figure~\ref{fig3} shows the results for the slopes as a function of the cell thickness $L$ for both procedures.
We observe that the variations are similar, but that the results arising from the two normalizations
differ by more than the error bars from the fit. Consequently, we calculate the  error bars
corresponding to the systematic effects from the normalization as being the half difference between the
results for the two procedures, for a given thickness.
In the main text we show the results corresponding to the second normalization procedure
that does not depend on any normalization slope.
The final error bars displayed in Figs.~\ref{fig3}(a,b) of the main text
are the quadratic sum of the systematic and statistical errors (extracted from a set of 5 measurements) as
well as of the errors from the fit.